\documentclass[11pt]{article}
\usepackage{amstext,amsfonts,amsmath,amssymb,amsthm}
\usepackage[all, knot]{xy}
\usepackage[usenames,dvipsnames]{color}

\newcommand{\reddish}{}
\newcommand{\blueish}{}
\newcommand{\pinkish}{}

\newfont{\gl}{eufm10 scaled \magstep1} %% gothic fonts
\newcommand{\dd}{\hbox{d}}

\def\Im{{\rm Im\,}}
\def\Re{{\rm Re\,}}
\def\ee{{\rm e}}

\newcommand{\beq}{\begin{equation}}
\newcommand{\eeq}{\end{equation}}

\newcommand{\beqs}{\begin{equation*}}
\newcommand{\eeqs}{\end{equation*}}

\numberwithin{equation}{section}
\begin{document}
\newcommand{\KeyssAndCodesMath}[2]{%
  \begin{quote} \small
   \textit{Keywords Mathematics:} #1.% 
   \def\@temp{#2}
   \ifx\@temp\@empty\relax  % <--- 3/11/06
   \else \par\textit{MSC2000:} #2.\fi
  \end{quote}} 
\newcommand{\KeyssAndCodesPhys}[2]{%
  \begin{quote} \small
   \textit{Keywords Physics:} #1.% 
   \def\@temp{#2}
   \ifx\@temp\@empty\relax  % <--- 3/11/06
   \else \par\textit{PACS numbers:} #2.\fi
  \end{quote}}

% ====================================================================
% Title, authors, abstract, keywords and AMS codes
% ====================================================================

\title{{Generalized Central Limit Theorem and Renormalization Group}}

\author{Iv\'an Calvo${}^{2,3}$\footnote{ivan.calvo@ciemat.es} , 
        Juan C. Cuch\'{\i}${}^{4}$\footnote{cuchi@eagrof.udl.cat} , 
        Jos\'e G. Esteve${}^{1,3}$\footnote{esteve@unizar.es} ,\\[3pt]
      and Fernando Falceto${}^{1,3}$
\footnote{Corresponding author. E-mail: falceto@unizar.es}\\[10pt]
${}^1$ Departamento de F\'{\i}sica Te\'orica, Universidad de Zaragoza,\\
50009 Zaragoza, Spain\\
${}^2$ Laboratorio Nacional de Fusi\'on, Asociaci\'on EURATOM-CIEMAT,\\
28040 Madrid, Spain\\
${}^3$ Instituto de Biocomputaci\'on y F\'{\i}sica de Sistemas
Complejos (BIFI),\\ 50009 Zaragoza, Spain\\
${}^4$ Departament d'Enginyeria Agroforestal, Universitat de Lleida, \\ 25198
  Lleida, Spain
}

\date{}

\maketitle

\begin{abstract}

\noindent
We introduce a simple instance of the renormalization group
transformation in the Banach space of probability densities. By
changing the scaling of the renormalized variables we obtain, as fixed
points of the transformation, the L\'evy strictly stable laws. We also
investigate the behavior of the transformation around these fixed
points and the domain of attraction for different values of the
scaling parameter. The physical interest of a renormalization group
approach to the generalized central limit theorem is discussed.

\end{abstract}
\KeyssAndCodesMath{Central Limit Theorems, Stable Distributions,
 Characteristic Functions}{60F05; 60E07; 60E10}
\KeyssAndCodesPhys{Renormalization Group Methods}{05.10.Cc}

\section{Introduction}

The classical Central Limit Theorem states that \reddish{the
  properly rescaled, centered} sum of $n$ independent and identically
distributed random variables with finite mean and variance converges
to the Gaussian distribution when $n\to\infty$. This universality
explains the ubiquity of the Gaussian distribution in the applications
of probability theory to many branches of science. In Physics, for
example, the Central Limit Theorem is behind such fundamental results
like the Maxwell-Boltzmann distribution and the microscopic
interpretation of diffusion by Einstein~\cite{Einstein}.

Diffusion equations are obtained as hydrodynamic limits of continuous
time random walks~\cite{MonWei} where the probability distribution of
the waiting-time is Markovian with mean waiting-time $\tau$, and the
probability distribution of the step-size has finite variance,
$\sigma$. Therefore, the microscopic transport mechanism has finite
characteristic length and time scales and the diffusion coefficient is
proportional to $\sigma^2/\tau$. However, in the last years and in
connection with the study of complex systems, it has been discovered
that many processes in Physics, Biology,
Economy \reddish{and the Social Sciences} exhibit scale-free
transport~\cite{Bouchaud,Metzler00,Zaslavsky02}, i.e. transport in
which characteristic spatial and/or temporal scales are lacking. The
so-called {\it anomalous diffusion} of these systems is understood in
terms of non-Gaussian statistics of the underlying microscopic
processes and modeled by means of fractional differential
equations~\cite{Fogedby,Podlubny}. The Generalized Central Limit
Theorem~\cite{Levy,GneKol} gives all the possible limits of sums of
(properly rescaled) independent and identically distributed random
variables, without the hypothesis of finite variance. Precisely the
limit distributions with infinite variance, usually called {\it L\'evy
distributions}, are the interesting ones for understanding scale-free
phenomena.

On the other hand, the Renormalization Group is the main theoretical
tool to investigate the universality that appears in different
branches of Mathe\-matics and Physics. Essentially, the Renormalization
Group explains how a system changes when the scale of observation is
modified. A Renormalization Group transformation usually consists in
averaging certain degrees of freedom in a way that the original system
is mapped to another with fewer degrees of freedom and different
coupling constants. Such transformation defines a flow in the space of
theories, with the fixed points and their linear stabi\-lity properties
giving much information about the large scale behavior of the system.

There is an intimate relationship between the
Central Limit
Theorem and the Renormalization Group. In particular, a proof of the
former can be given from the perspective of the latter. This is
helpful in order to understand in a different way the mechanism of
convergence of the sums of distributions, which are viewed as
iterations of certain Renormalization Group transformations. This
reformulation was rigorously done by G. Jona--Lasinio~\cite{Jona1} for
the classical Central Limit Theorem, see also~\cite{Honkonen,Jona2,KorSin}.

As far as we know, the aforementioned appearance of L\'evy statistics
in a number of transport processes is not yet well-understood from
first principles. Let us focus for a moment in the observed anomalous
transport in certain regimes of turbulent fusion
plasmas~\cite{delCastillo05,Mier08,Sanchez08}. The conundrum can be
informally posed as follows: the fundamental equations are non-linear
partial differential equations (fluid momentum balance equation
coupled to Maxwell equations, for example), but particle transport
seems to be suitably modeled in terms of linear fractional
differential equations. Let us simply mention that whereas the symbol
in Fourier space of an ordinary derivative operator is $(-ik)^n$ with
$n\in{\mathbb Z}$, the symbol of a fractional derivative operator is
$(-ik)^\alpha$ with $\alpha\in{\mathbb R}$. A deep understanding of
this change in the analyticity properties of the involved operators
(or equivalently, a deep understanding of the emergence of L\'evy
statistics) is still lacking. From the point of view of Physics it is
natural to explore the application of Renormalization Group
ideas. Although a satisfactory and complete answer is probably far
ahead, in the present paper we try to take a first step by showing how
the Generalized Central Limit Theo\-rem and therefore L\'evy
distributions show up from Renormalization Group arguments. We put
special emphasis on the study of the flow in the space of probability
distributions. We think that this might open a way to make further
progress in the comprehension of the problems stated above.

We find it valuable to further motivate a Renormalization Group
approach to the Generalized Central Limit theorem by working out
rather briefly a simple but interesting physical model in which such a
perspective emerges. Consider a discrete-time random walk equation
\begin{equation}\label{ranwalk}
n(x,t+\tau)=\int\rho(x-y)n(y,t){\rm d}y,
\end{equation}
where $n(x,t)$ is the density of walkers and $\rho$ is an arbitrary
symmetric pro\-ba\-bility density function (p.d.f. in the following),
so that $\rho(u){\rm d}u$ 
is the probability of a step taking value in 
$[u, u+{\rm d}u]$. 
Jumps take place at discrete-time intervals of length $\tau$,
i.e. $t=n\tau$, $n\in{\mathbb N}$. The so-called fluid or hydrodynamic
limit of this equation is obtained by considering that the previous
process occurs at a microscopic scale \reddish{well-separated
  from} the macroscopic observational space-time scales. There are at
least two ways of implementing this idea.

The first way (see \cite{Scalas04} for a clear and detailed treatment
in more general random walk models) consists in relating the
microscopic and macroscopic characteristic time and length by means of
a scaling factor $\lambda$ that eventually goes to infinity. As long
as the dimensions are properly chosen, (\ref{ranwalk}) yields a
(fractional) differential equation when $\lambda\to\infty$. Namely,
define

$$\reddish{{\mathbf{n}(\mathbf{x},\mathbf{t}):=} n(\lambda^\Delta\mathbf{x},
\lambda\mathbf{t}),}$$
the density in macroscopic units, where $\lambda$ is the scaling
factor that converts the microscopic scale into the macroscopic one.
The dimension $\Delta$ has to be adjusted to obtain a non trivial
fluid limit when $\lambda\to\infty$. We can write now the
microscopic random walk equation (\ref{ranwalk}) in terms of the
macroscopic density
\begin{equation}
\mathbf{n}(\mathbf{x},\mathbf{t}+\tau/\lambda)=\lambda^{\Delta}\int\rho(\lambda^{\Delta}(\mathbf{x}
-\mathbf{y}))\mathbf{n}(\mathbf{y},\mathbf{t}){\rm d}\mathbf{y}.
\end{equation}
Or considering its Fourier transform
\begin{equation}
\hat{\mathbf{n}}(\mathbf{k},\mathbf{t}+\tau/\lambda)=
\hat\rho(\lambda^{-\Delta}\mathbf{k})\hat{\mathbf{n}}(\mathbf{k},\mathbf{t}).
\end{equation}
If we assume
\begin{equation}
\hat\rho(k)=1-c|k|^\alpha+o(|k|^\alpha),
\end{equation}
for some $\alpha\in(0,2]$, $c>0$, and take $\Delta=1/\alpha$, we
  obtain a non-trivial limit when $\lambda\to\infty$:
\begin{equation}
\partial_{\mathbf{t}}\hat{\mathbf{n}}(\mathbf{k},\mathbf{t})
=-\frac{c}\tau|\mathbf{k}|^\alpha
\hat{\mathbf{n}}(\mathbf{k},\mathbf{t}),
\end{equation}
that represents a diffusion equation with fractional derivatives for
the density function $\mathbf{n}(\mathbf{x},\mathbf{t})$. In
particular, Einstein's derivation of the diffusion equation from a
random walk model corresponds to $\alpha=2$.

Note that in the previous limit most of the details of the microscopic
process have disappeared and only the lowest order in the expansion of
$\hat\rho-1$ around $k=0$ is relevant for determining the form of the
fractional diffusion equation.  This is a manifestation of
universality, typical in phenomena that emerge in the macroscopic
domain. Observe also that the description obtained at the macroscopic
level, a fractional differential equation, is of different nature to
that at the microscopic level, a discrete time random walk equation.

The second way to tackle the problem is inspired by the
Renormalization Group ideas. We will see that in this second approach
the same description of the process, i.e. a discrete time random walk,
is kept in the microscopic and macroscopic regimes. This procedure is
specially useful in situations like quantum field theory where, due to
the appearance of infinities, a cut-off is usually
\reddish{required} to define the theory. To understand how this
works we start by iterating (\ref{ranwalk}) to get
\begin{equation}\label{ranwalk2}
n(x,t+2\tau)=\int\rho*\rho(x-y)n(y,t){\rm d}y,
\end{equation}
where $\rho*\rho(x)=\int\rho(x-y)\rho(y)dy$
stands for the convolution of $\rho$ with itself.

In order to compare this equation with the original one we perform a 
rescaling in space and time and introduce the new density
\reddish{
$$n'(x',t')=n(ax',2t'), \quad a>0.$$
}
In terms of this density \reddish{Eq. (\ref{ranwalk2})} reads
\begin{equation}\label{ranwalkren}
n'(x',t'+\tau)=\int T_a\rho(x'-y')n'(y',t'){\rm d}y',
\end{equation}
that looks exactly like (\ref{ranwalk}) except that the
\reddish{p.d.f.} has changed to
$$T_a\rho(x)=a\rho*\rho(ax).$$
We call $T_a$ the renormalization group transformation.

If we examine carefully what (\ref{ranwalkren}) means, we observe that
although it looks like a process in discrete time $\tau$ with
\reddish{p.d.f.} $T_a\rho$, it actually represents the original
  process with \reddish{p.d.f.} $\rho$, discrete time $\tau/2$
  and rescaled space variable.  The fluid or diffusive limit is
  obtained when we apply this procedure infinitely many times, so that
  we can write
\begin{equation}
\mathbf{n}(\mathbf{x},\mathbf{t}+\tau)=\int\boldsymbol{\rho}
(\mathbf{x}-\mathbf{y})\mathbf{n}(\mathbf{y},\mathbf{t}){\rm d}\mathbf{y}.
\end{equation}
where 
$$\boldsymbol{\rho}=\lim_{m\to\infty}T_a^m\rho.$$ Therefore, in order
to have a well defined process in the fluid limit we should adjust $a$
so that the previous limit exists.  $\boldsymbol{\rho}$ is necessarily
a fixed point of the renormalization group
transformation. Universality in this approach shows up because the
fluid limit is the same for all initial \reddish{p.d.f.'s}
$\rho$ that belong to the domain of attraction of the same fixed point
under the action of $T_a$. All these considerations motivate the
interest of the study of the renormalization group transformation
$T_a$, its fixed points and their stability.

The rest of the paper is organized as follows. In Section 2 we
introduce more precisely the renormalization group transformation and
study the behavior of the moments.  In Section 3 we establish the
appropriate differential setup for investigating topological
properties of the transformation. Finally, Section 4 is devoted to the
study of the fixed points of the transformation, their stability, and
their domain of attraction.

\section{Renormalization group transformation}
Let $\rho$ be a {p.d.f.}  
 in ${\mathbb R}$, i.e.
\begin{itemize}
\item[(i)] $\rho(x)\ge 0$, $\forall x\in \mathbb{R}$ \ \ and
\item[(ii)] $\int_{-\infty}^\infty\rho(x){\rm d}x=1$;
\end{itemize}
we define the following {Renormalization Group} transformation acting on it
\begin{equation*}
T_{a}\rho (x) :=|{a}|
\int_{-\infty}^\infty\rho({a} x-s)\rho(s)\dd s,
\end{equation*}
where
${a}$ is a real number different from zero. Note that
for normalized $\rho$,  $T_{a}\rho$ is also normalized.

The meaning of the transformation is clear; $T_{a}\rho$ is the probability
density of the random variable $\xi'=(\xi_1+\xi_2)/{a}$
where $\xi_1, \xi_2$ are independent identically distributed random
variables with density $\rho$.
The two essential ingredients of the renormalization group:
new variables that represent an average of the old ones and
the scaling of the former,
are present in this example in the simplest way.

To understand how the different values of ${a}$ lead to different
properties of the transformation it is instructive to consider the
situation in which all moments
\begin{equation}
\langle x^n\rangle_\rho=\int_{-\infty}^\infty x^n \rho(x) \dd x,
\quad n\in \mathbb{N},
\end{equation}
are finite (in the subsequent sections we will work out the general
 case) and study their behavior under the transformation. One
 immediately obtains the following expression for the moments of the
 transformed probability distribution:
\begin{equation*}
\langle x^n\rangle_{T_{a}\rho}={a}^{-n}\sum_{i=0}^n
\binom{n}{i}
\langle x^i\rangle_\rho\langle x^{n-i}\rangle_\rho,
\quad n\in \mathbb{N}.
\end{equation*}

If we take in particular $n=1$ we get:
\begin{equation*}
\langle x\rangle_{T_{a}\rho}=\frac{2}{a}\langle x\rangle_\rho,
\end{equation*}
which implies that the transformation maps densities with zero mean
into themselves.  \reddish{In this case the relation} between
the variances is
\begin{equation*}
\langle x^2\rangle_{T_{a}\rho}=
\frac{2}{{a}^2}\langle x^2\rangle_\rho.
\end{equation*}

Now, we can distinguish three different regimes after $m$ successive
applications of the transformation $T_a$ according to the values of
${a}$, when $m\to\infty$:
\begin{itemize}
 \item[--] For $|{a}|>\sqrt{2}$,
       \begin{equation*}
       \lim_{m\to\infty}\langle x^2\rangle_{T_{a}^m\rho}=0,
       \end{equation*}
       which implies that for distributions with finite variance,
       $T^m_{a}\rho$ should approach the Dirac delta function
       in the large $m$ limit. In the next section we
       shall make precise the topological space in which $T_{a}$ acts
       and we shall see in which sense the above limit holds.

 \item[--] For $|{a}|<\sqrt{2}$, assuming that $\langle x^2\rangle_\rho \not= 0$, we have
           \begin{equation*}
           \lim_{m\to\infty}
           \langle x^2\rangle_{T_{a}^m\rho}=\infty.
           \end{equation*}

 \item[--] Finally, when  $|{a}|=\sqrt{2}$, the value of $\langle x^2 \rangle$ does
           not change under the application of the renormalization group
           transformation.  In this case we have a fixed point with finite
           non-zero moments. Denoting by $\rho_{_0}$ the fixed point, $T_{a}
           \rho_{_0}=\rho_{_0}$, we will have that
           \begin{equation*}
           \langle x^{2n}\rangle_{\rho_{_0}} =\frac{1}{2^n-2}\sum_{i=1}^{n-1}
           \binom{2n}{i}
           \langle x^{2i}\rangle_{\rho_{_0}}
           \langle x^{2(n-i)}\rangle_{\rho_{_0}},
           \end{equation*}
           which is solved by
           \begin{equation*}
           \langle x^{2n}\rangle_{\rho_{_0}}=
           (\langle x^{2}\rangle_{\rho_{_0}})^n\frac{(2n)!}{ n! 2^n}.
           \end{equation*}
           That solution coincides, of course, with the expression for the moments of the Gaussian
           distribution.
\end{itemize}

\section{Banach space structure}

The natural framework to ask topological and differential questions
about $T_{a}$ is that of Banach spaces.  In our case we shall consider
test functions in $C_\infty({\mathbb R})$: the Banach space of
continuous functions with vanishing limit at $\infty$, endowed with
the supremum norm. The probability densities are the positive
distributions of unit norm in its topological dual $C_\infty({\mathbb
  R})'$.  The space $C_\infty({\mathbb R})'$ consists of the
\reddish{finite}, complex Radon measures in $\mathbb R$; it
  contains, for instance, distributions supported on discrete sets
  (Dirac delta functions) as well as densities in $L^1({\mathbb R})$.

We shall denote the action of an element of
$\rho\in C_\infty({\mathbb R})'$ on
$f\in C_\infty({\mathbb R})$, $\rho f$, by
\begin{equation*}
\rho f:=\int_{-\infty}^\infty\rho(x)f(x){\rm d}x.
\end{equation*}

It is a classical result, see \cite{Rudin} for instance, that the
convolution $\rho_1*\rho_2$ of two finite Radon measures is again a
finite Radon measure and actually $C_\infty({\mathbb R})'$ is a
commutative Banach algebra with the convolution.  \reddish{In
  particular, the convolution is associative and satisfies}
\begin{equation*}
\parallel\rho_1*\rho_2\parallel\leq\parallel\rho_1\parallel\,\parallel\rho_2\parallel,
\end{equation*}
where the equality is attained if both distributions are positive.

To define the renormalization group transformation in
$C_\infty({\mathbb R})'$ we introduce the {\it dilation}
$\pi_r:C_\infty(\mathbb R)\rightarrow C_\infty(\mathbb R)$,
\begin{equation*}
(\pi_r f)(x):=f(r x).
\end{equation*}
Now we can obtain the transformed distribution under the renormalization
group by means of the formula
\begin{equation}\label{transformation}
T_{a} \rho= (\rho *\rho)\pi_{1/{a}}
\end{equation}
that correctly defines a linear functional
$T_{a} \rho: C_\infty({\mathbb R})\rightarrow {\mathbb R}$.

The main differential properties of the renormalization group 
transformation are collected in the following proposition.
\vskip .5cm
\noindent{\bf Proposition 1:}
\vskip .1cm
{\it The transformation $T_a: C_\infty({\mathbb R})'\rightarrow C_\infty({\mathbb R})'$ in
(\ref{transformation}) is continuous and differentiable
with bounded continuous differential.}
\vskip .3cm
\noindent{\it Proof:}  
\noindent To show the continuity of $T_a$ it is enough to realize that
$\pi_{1/{a}}$ is an isometry. Then, we get
\begin{align}
\parallel T_{a} \rho_1- T_{a} \rho_2\parallel&=
\parallel(\rho_1+\rho_2)*(\rho_1-\rho_2)\pi_{1/{a}}\parallel
\notag
\\
&\leq\parallel\rho_1+\rho_2\parallel\,\parallel\rho_1-\rho_2\parallel,
\end{align}
from which the continuity of $T_{a}$ follows. On the other hand, the
differential of $T_{a}$ is given by
\begin{equation}\label{differential}
(DT_{a})_\rho\zeta=2(\rho *\zeta)\pi_{1/{a}},
\end{equation}
which is a bounded linear map. In fact 
one has
\begin{align}
\parallel (DT_{a})_\rho\zeta\parallel&=
2\parallel \rho*\zeta \parallel\leq 2\parallel \rho\parallel
\,\parallel\zeta\parallel.
\notag
\end{align}
Therefore,
\begin{equation*}
\parallel(DT_{a})_\rho\parallel\leq 2\parallel\rho\parallel,
\end{equation*}
and the equality is attained if $\rho$ is positive.
As for the continuity of the differential simply observe that
\begin{equation*}
\parallel(DT_{a})_{\rho_1}-(DT_{a})_{\rho_2}\parallel
=
\parallel(DT_{a})_{\rho_1-\rho_2}\parallel\leq2\parallel\rho_1-\rho_2\parallel,
\end{equation*}
and the proof is concluded.\qed
 
\section{Fixed points}

Much information about the renormalization group transformation is
obtained by studying its fixed points and its linearization around
them. In order to work out the fixed points it is convenient to
express the transformation in terms of the characteristic functions
\begin{equation*}
\widehat{T_{a}\rho}(k)=\int{T_{a}\rho(x)\ee^{ikx} dx}= \widehat\rho(k/{a})^2.
\end{equation*}
Then, for the fixed point
$T_{a}\rho_{_0}=\rho_{_0}$, one has
\begin{equation}\label{fixpointk}
\widehat\rho_{_0}(k/{a})^2=\widehat\rho_{_0}(k).
\end{equation}
Note that this equation is solved by functions $\widehat\rho$
whose logarithm is a symme\-tric, homogeneous function
of degree $\alpha=\log2/\log|a|$.
To find more general solutions
we distinguish two cases according to the sign of $a$.
\begin{itemize}
 \item [--] If $a$ is positive, we have solutions for 
            (\ref{fixpointk}) of the form
            \begin{equation*}
            \widehat\rho_{_0}(k)=\widehat S_{\alpha,A}(k):=
            \exp(-A\vert k\vert^\alpha\theta(k)-\overline{A}\vert
            k\vert^\alpha\theta(-k)),
            \end{equation*}
            where $\alpha$ is given by the relation $\vert
            a\vert=2^{1/\alpha}$, $\overline{A}$ is the complex
            conjugate of $A$ and $\theta$ is the Heaviside step
            function.
            
            The fact that $\widehat S_{\alpha,A}$ is the
            characteristic function of a probability density (i.e. a
            positive functional) imposes additional restrictions on
            $A$ and $\alpha$. In particular it is well-known (see
            \reddish{Refs. \cite{Levy,GneKol}} for instance) that if
            $A\not=0$, then we must have
            \begin{equation}\label{conditions}
            0<\alpha\leq 2\quad{\rm and}\quad A =\vert
            A\vert\ee^{i\varphi}\quad {\rm with}\quad
            \vert\varphi\vert \leq\frac\pi2(1-\vert \alpha-1\vert).
            \end{equation}
            The solution with $A=0$ defines a fixed point of the
renormalization group, corresponding to the Dirac delta function, that
exists for any value of $a$.

            These solutions of (\ref{fixpointk}) are the so-called
            strictly stable laws, denoted here by $S_{\alpha,A}$, that
            where introduced by \reddish{P. L\'evy}~\cite{Levy}, see
            also~\cite{GneKol,IbrLin,AarDen}.

 \item [--] For negative $a$, the positive and negative values
            of $k$ are related through (\ref{fixpointk}). As a result,
            the fixed points are the ones described above for $\vert a\vert$
            but restricted to real values of $A$.
\end{itemize}

After discussing the fixed points of our renormalization group
transformation we analyze their stability and the domain of attraction
(around the fixed point). The standard way to approach this problem is
to study the spectrum and eigenvectors of the differential of the
transformation at the fixed points.  The results can be summarized in
the following
\vskip .3cm
\noindent{\bf Proposition 2:}\pinkish{
\vskip .1cm {\it Let $(DT_{a})_\rho$ be the operator defined in
(\ref{differential}), $\rho_0$ one of the fixed points of $T_a$
discussed above, and ${\boldsymbol\sigma}$ the complex spectrum of
$(DT_{a})_{\rho_0}$.
\begin{itemize}
\item [(i)] If $\rho_{_0}=S_{\alpha,A}$
%$\widehat\rho_{_0}(k)= \exp(-A\vert
%k\vert^\alpha\theta(k)-\overline{A}\vert k\vert^\alpha\theta(-k))$
with 
${\Re} A > 0$, 
then $\boldsymbol{\sigma}=\{\lambda,\;\; \text{s. t.}\;\; |\lambda|\leq2\}$
and all its values belong to the pointwise spectrum except for
$\lambda=0$.
\item [(ii)] If 
$\rho_{_0}=S_{\alpha,A}$ with ${\Re} A = 0$, 
then $\boldsymbol{\sigma}=\{\lambda,\;\; \text{s. t.}\;\;
\vert\lambda\vert=2\}$ with its only eigenvalue at $\lambda=2$.
\end{itemize}
}}
\vskip .5cm
\noindent{\it Proof:} In order to determine the pointwise spectrum one
has to solve the eigenvalue equation
\begin{equation}\label{eigenvalues}
(DT_a)_{\rho_{_0}}\zeta=\lambda\zeta.
\end{equation}

We will focus first on (i), corresponding to the non-trivial fixed
point $\rho_0=S_{\alpha,A}$ with \blueish{${\Re}A>0$}.  The
equation for the eigenvalues (\ref{eigenvalues}) can be more easily
handled by writing it in terms of the characteristic
\reddish{functions, $\widehat\zeta$. Namely,}
$$2 {\widehat S}_{\alpha,A}(k/\alpha)\widehat\zeta(k/\alpha)=\lambda
\widehat\zeta(k).$$

\reddish{Using the {\it ansatz}}
$\widehat\zeta(k)={\widehat S}_{\alpha,A}(k)\widehat\eta(k)$
and taking into account the properties of the fixed point, 
we arrive at the following equation 
\begin{equation}\label{eigeneta}
2\widehat\eta(k/a)=\lambda\widehat\eta(k),
\end{equation}
which can be solved with homogeneous functions
of the appropriate degree. 
Namely, for positive $a=2^{1/\alpha}$ 
we have solutions of the form
\begin{equation}
\blueish{
\widehat\eta_s(k)=(B_+\theta(k)+B_-\theta(-k))|k|^s
}\end{equation}
with eigenvalue $\lambda_s=2^{1-\frac{s}\alpha}$.

In order to ensure that $\zeta=S_{\alpha,A}*\eta\in C_\infty({\mathbb
  R})'$ we must have ${\Re} s \ge 0$\reddish{. We}
\pinkish{shall now show} that it is also
sufficient. \reddish{The function}
$$\zeta_s(x)=\frac1{2\pi}\int_{-\infty}^\infty  
\blueish{\widehat\eta_s(k)}
\ee^{-A\vert
k\vert^\alpha\theta(k)-\overline{A}\vert k\vert^\alpha\theta(-k)} 
\ee^{-ikx}\ \dd k,
$$ \blueish{where ${\Re}A>0$ and ${\Re} s \ge 0$,}
\reddish{is clearly continous and even smooth, so} we only need
to show that it decays sufficiently fast for large $|x|$. Consider the
integration of the positive momenta

$$\zeta_{_+}(x)= \frac{
\blueish{B_+}
}{2\pi}\int_{0}^\infty k^s
 \ee^{-A k^\alpha-ikx} \dd k.
$$

Assuming that $x>0$ and \blueish{${\Re}A>0$}
we can perform a slight clockwise rotation of 
angle $\gamma$ of the integration line in the complex plane. 
It does not affect 
the result due to \reddish{Cauchy's theorem}. 
If we also change to a new variable
$z=\ee^{i\gamma}kx$
the integral reads 
$$\zeta_{_+}(x)
=\frac{
\blueish{B_+}
\ee^{-i\gamma(s+1)}}{2\pi x^{s+1}}\int_{0}^\infty z^s
 \ee^{-A (z/x)^\alpha \ee^{-i\alpha\gamma}-iz\ee^{-i\gamma}} \dd z
$$
and for small $\gamma > 0$ we can safely take the limit of large $x$
inside the integral to get
$$
\zeta_{_+}(x)= \frac{f(x)}{x^{s+1}}, \qquad {\rm with}\quad 
\lim_{x\to \infty}f(x)=\blueish{B_+}
\frac{\Gamma(s+1)}{2\pi}\ee^{-i\pi(s+1)/2}.
$$
We can deal in an analogous way with the integration on the negative semiaxis
and also with the case of negative $x$. In all cases we can show that
provided ${\Re}s>0$ the function decays fast enough so that
$\zeta=S_{\alpha,A}*\eta\in C_\infty({\mathbb R})'$ 

In this way we prove that all values of $\lambda$ s.t.
$0<|\lambda|\leq 2$ are in the pointwise spectrum. As the spectrum
must be a closed set, $0$ is also in $\boldsymbol\sigma$.  To exclude
other points, i.e $|\lambda|>2$ we can use the well-known argument
that if $|\lambda|> \parallel (DT_a)_{\rho_0}\parallel =2$ then it
belongs to the resolvent. \pinkish{This concludes the
  proof of part (i) of the proposition}.

\pinkish{We proceed to prove part (ii).  There are
  only two possibilities of having a fixed point $S_{\alpha, A}$ with
  ${\Re}A=0$: either $\alpha=1$ and $\varphi=\pm\pi/2$ or $A=0$. In
  all other instances ${\Re}A>0$.  We shall discuss in detail the
  fixed point with $A=0$, i.e. the Dirac delta function $\delta$,
  which exists for any value of $a$}.  To study the spectrum of the
differential at this point we introduce the operator
\begin{equation*}
U\zeta:=\frac12(DT_a)_\delta\zeta=\zeta\pi_{1/a}.
\end{equation*}
$U$ is an isometry and, therefore, its spectrum lies entirely in 
the unit circle. On the other hand the equation for the eigenvalues
\begin{equation}\label{eigenvalues2}
U\zeta_\mu=\mu\zeta_\mu
\end{equation}
is solved with $\mu=1$ and $\zeta_{_1}=\delta$.  To exclude the
possibility of any other solution we can use the following simple
argument (valid for $a=2^{1/\alpha}>1$, although a minor modification
allows to deal with the case $a<-1$).  

Consider a function $\phi\in C_\infty({\mathbb R})$ with
$\phi(a)=\bar\mu\phi(1)$ and $\parallel \phi\parallel=1$, but
arbitrary \reddish{otherwise. Now} for any positive integer $N$
we construct the function $f^{(N)}\in C_\infty({\mathbb R})$ defined
for $x\in[a^{-N},a^N]$ by
$$f^{(N)}(a^n x)=\bar\mu^n\phi(x), \quad x\in[1,a], \quad
n=-N,-N+1,\dots,N-1,N,$$ and extended \reddish{to $\mathbb R$}
so that $\parallel f^{(N)}\parallel=1$. \reddish{Thus,}
\begin{align}
&&\int_{wa^n}^{wa^{n+1}}\zeta_\mu(x) f^{(N)}(x)\ \dd x=
\int_{wa^{n+1}}^{wa^{n+2}}\zeta_\mu(x) f^{(N)}(x)\ \dd x=I_w,\cr
&&\qquad w\in(1,a),\quad n=-N,\dots,N-3.
\end{align}
Then
$$
\int_{wa^{-N}}^{wa^{N-1}}\zeta_\mu(x) f^{(N)}(x)\dd x=(2N-1)I_w
$$ and, if $I_w\not=0$, $\zeta_\mu$ is \blueish{an unbounded
  functional}. Therefore $I_w=0$, \pinkish{implying
  that the intersection of the support of $\zeta_\mu$ with
  $(0,\infty)$ is empty}. A similar argument rules out the negative
semiaxis and finally one deduces that $\zeta_\mu$ must be supported
\reddish{in $\{0\}$}, i.e. it is the $\delta$-function that
corresponds to $\mu=1$.

As for the rest of the spectrum consider, for $\mu=\ee^{i\theta}$ and 
a positive integer $\blueish{K}$, the following 
distribution $\zeta_\mu^{(\blueish{K})}\in C_\infty'(\mathbb{R})$:
\begin{equation} 
\zeta_\mu^{(\blueish{K})}(x)=
\begin{cases}
0& {\rm if}\ |x|\not\in(1/\blueish{K},\blueish{K}),\\
\displaystyle\frac1{4\log \blueish{K}}
 \frac{\ee^{i\theta\log|x|/\log|a|}}{|x|}& {\rm if}\ 
|x|\in(1/\blueish{K},\blueish{K}).
\end{cases}
\end{equation}
One can compute $\parallel
\zeta_\mu^{(\blueish{K})}\parallel=1$ and

$$\reddish{\parallel
  U\zeta_\mu^{(\blueish{K})}-\mu\zeta_\mu^{(\blueish{K})}\parallel=
  \frac{\log a}{\log \blueish{K}}\, ,}$$ which shows that
$\mu=\ee^{i\theta}$ is in the spectrum of $U$. Therefore, the complex
spectrum of $(DT_a)_\delta=2U$ is as stated in the second part of the
proposition. 

\pinkish{As mentioned before, the only other
   possibility to obtain a fixed point with ${\Re}A=0$ is having
   $\alpha=1$ and $\varphi=\pm\pi/2$.} \blueish{The fixed point
     in this case is $\delta(x+e)$ with $e=\Im A$. The proof above can
     be applied to this case by simply shifting to $-e$ the origin of
     the real line.}  \qed
\vskip .5cm

We would like to discuss now the information about the behavior of the
renormalization group transformation around the fixed points that can
be extracted from the study of the spectrum. We start by considering
the fixed points corresponding to strictly stable L\'evy
distributions.  In this case the stable directions that lie on the
domain of attraction of the fixed point correspond to $|\lambda_s|<1$,
i.e. $s>\alpha$, while those with $s<\alpha$ \reddish{give
  $|\lambda_s|>1$} and therefore unstable perturbations of the fixed
  point.

It is interesting to analyze the behavior
of the characteristic function of the
fixed point perturbed in the \blueish{$\zeta_s$}
direction for small values of $k$.
First note that
\begin{equation*}
\widehat\rho_{_0}(k)
= 1-A\vert k\vert^\alpha\theta(k)-
\overline{A}\vert k\vert^\alpha\theta(-k)+ o(|k|^\alpha)
\end{equation*}
and that
\begin{equation*}
\widehat\rho_{_0}(k)+\epsilon\widehat\zeta_s(k)=
\begin{cases}
1-A\vert k\vert^\alpha\theta(k)-\overline{A}\vert k\vert^\alpha\theta(-k)+o(|k|^\alpha)&{\rm for}\
s>\alpha,\cr
1+ B_+k^s\theta(k)+ B_-|k|^s\theta(-k)+o(|k|^s)& {\rm for}\ s<\alpha,
\end{cases}
\end{equation*}
which implies that the perturbations that do not change the
behavior of the characteristic function around $k=0$ are precisely
the stable directions.  In this way we obtain an infinitesimal
version of the well-known result~\cite{GneKor} that the
the p.d.f.s whose characteristic function behaves around $k=0$ as
\begin{equation*}
\widehat\rho(k)=1-A\vert k\vert^\alpha\theta(k)-\overline{A}\vert
k\vert^\alpha\theta(-k)+o(|k|^\alpha)
\end{equation*}
belong to the domain of attraction of the fixed point
with characteristic function
\begin{equation*}
\widehat\rho_{_0}(k)=\exp(-A\vert k\vert^\alpha\theta(k)
-\overline{A}\vert k\vert^\alpha\theta(-k)).
\end{equation*}

We consider now the other fixed point, the Dirac delta function. From
the results of Section 2 (see also below) one would expect to have an
attractive fixed point at $\delta$, at least for certain values of
$a$. Therefore, it may seem surprising that the spectrum of the
differential around the fixed point $\delta$ has only eigenvalues with
modulus greater than 1. The reason is that the eigenvectors with
$|\lambda|<1$ are associated to the strong convergence of the iterated
transformation of the fixed point perturbed in that direction, while
in the case of $\delta$, strong convergence is impossible and we have
convergence only in the weak topology. This fact can be shown by
looking more closely at the behavior of the transformation around the
fixed point.  Consider a perturbation of $\delta$ by $h\in
C'_\infty({\mathbb R})$ with $0<\parallel h \parallel<1$. Then
\begin{align}
\parallel T_a(\delta+h)-&\delta\parallel = \parallel 2h+h*h\parallel\geq\hfill\cr
&\geq 2\parallel h\parallel -\parallel h*h\parallel 
\geq 2\parallel h \parallel-\parallel h \parallel^2 > \parallel h \parallel,
\end{align}
and all directions around the fixed point $\delta$ are unstable in the
strong topology.  The situation in the weak topology is different and,
\reddish{as shown below}, there are p.d.f.s in $L^1$
\reddish{which under successive application of the renormalization
  group transformation converge} weakly to the Dirac delta function.

In order to complete the local picture obtained above and to substantiate
previous statements, we shall discuss
the behavior of the iteration of the transformation
for different initial densities and different values of $a$. Let
\begin{equation}\label{distribution}
\widehat\rho(k)=1-B|k|^\nu\theta(k)-
\overline B|k|^\nu\theta(-k)+o(|k|^\nu),
\end{equation}
with $0<\nu\leq2$ and $B\not=0$.
Therefore, for fixed $k$ and $a=2^{1/\alpha}>1$ one has
\begin{equation*}
\widehat{T^{^{\scriptstyle n}}_a\rho}(k)=\left(1-a^{-\nu n}B|k|^\nu\theta(k)-
a^{-\nu n}\overline B|k|^\nu\theta(-k)+o(a^{-\nu n})\right)^{2^{n}}
\end{equation*}
and the limit of large $n$ with $k\not=0$ reads
\begin{equation}\label{convergence}
\lim_{n\to\infty}\widehat{T^{^{\scriptstyle n}}_a\rho}(k)=
\begin{cases}
1&{\rm for}\ \nu>\alpha,\cr
\exp(-B|k|^\nu\theta(k)-
\overline B|k|^\nu\theta(-k))&{\rm for}\ \nu=\alpha,\cr
0&{\rm for}\ \nu<\alpha.
\end{cases}
\end{equation}

We therefore have the following proposition:
\vskip 0.5cm
{\bf Proposition 3:}
\vskip 0.1cm {\it Take $\rho\in C'_\infty({\mathbb R})$ with
characteristic function given by (\ref{distribution}) and
$a=2^{1/\alpha}$. Then, $T^n_a\rho$ converges weakly when $n\to\infty$
to $\delta$ if $\nu>\alpha$, to $S_{\alpha,B}$ if $\nu=\alpha$ and it
does not converge if $\nu<\alpha$.}
\vskip 0.3cm
\noindent The result is a simple consequence of L\'evy's continuity theorem
together with the behavior of the characteristic functions in (\ref{convergence}).\qed

\vskip 0.5cm
With a similar computation we obtain for $a=-2^{1/\alpha}<-1$ the
 {following:}
\begin{itemize}
 \item [--] If $\nu>\alpha$, $T^{^{\scriptstyle n}}_a\rho$
            converges weakly to the $\delta$ function.
 \item [--] For $\nu=\alpha$ we have a limit
            two-cycle formed by the strictly stable densities $S_{\alpha,B}$ 
            and $S_{\alpha,\bar B}$.
            The two members of the cycle are
            related by reflection with respect to the origin $x\mapsto-x$.
 \item [--] Finally, for $\nu<\alpha$ the sequence does not converge.
\end{itemize}

\vskip 0.5cm In order to illustrate some of the advantages of our
approach to the Ge\-ne\-ralized Central Limit Theorem we will give now a
simple proof, based on the above results, of the positivity of
$S_{\alpha,A}$ for $\alpha$ and $A$ as in (\ref{conditions}). The
proof goes as follows:

\noindent Take the following p.d.f.
$$\rho(x)=
{\frac{c_1}{n^{\alpha+1}+\vert x\vert^{\alpha+1}}}\theta(-x)+
{\frac{c_2}{n^{\alpha+1}+\vert x\vert^{\alpha+1}}}\theta(x),$$
with $c_i\geq0, c_1+c_2>0, \alpha\in(0,1)$. $n$ is a positive real number,
chosen so that $\parallel\rho\parallel=1$. Standard estimates show that
$$\widehat\rho(k)=1-A\vert k\vert^\alpha\theta(k)-\overline{A}\vert
k\vert^\alpha\theta(-k)+o(|k|^\alpha),
$$
where $A=|A|\ee^{i\phi}$ with 
$$|A|=(c_1^2+c_2^2+2c_1c_2\cos(\pi\alpha))^{1/2}
{\frac{\Gamma(1-\alpha)}\alpha}$$ 
and 
$$\tan\phi=\frac{c_2-c_1}{c_1+c_2}\tan(\frac{\pi\alpha}2)\quad {\rm
with}\ |\phi|\le\frac\pi2.$$
On the other hand, from our previous
results we \reddish{have}
$$\lim_{n\to\infty}T_a^n\rho=S_{\alpha,A},$$
which, due to the fact that our transformation 
preserves positivity, implies that $S_{\alpha,A}$ is positive.
Note that taking different non-negative values for $c_1$ and $c_2$ we 
cover the whole range for $A$ as described in (\ref{conditions}).

A slight modification of the initial p.d.f. and the estimate allows to
deal with the remaining cases, i.e. $\alpha\in[1,2]$.
 
\section{Conclusions}

In this paper we have studied a simple instance of the renormalization
group transformation on the space of probability densities. We have
shown that {by} changing the scaling that governs the
transformation {one obtains} different fixed points or
limit two-cycles that correspond to the strictly stable laws
 of {L\'evy} and the limiting case of the Dirac
 $\delta$ {distribution}.

We have also studied the stability of the fixed points using the
linear approximation of the {transformation} around the
fixed point. In this way we {have derived} 
{a local version of}
the classical results about the domain of {attraction} of
the stable laws.  We {have} also {shown}
that the stable or unstable character of the fixed point can be
different if we consider the strong or the weak topology in the space
of probability densities.

\vskip0.5cm

\reddish{     
\noindent{\bf Acknowledgements:} \blueish{We wish to thank the referees 
for a careful reading and useful comments that helped us to improve 
the paper.} I. C. and J. C. C. gratefully acknowledge the
hospitality of the Department of Theoretical Physics at
the University of Zaragoza, where part of this work was
done. Research partially supported by
grant FIS2006-01225 MEC (Spain), and grants
ENE2009-07247 and FPA2009-09638, 
Ministerio de Ciencia e Innovaci\'on (Spain).
}

%%%%%%%%%%%%%%%%%%%%%%%%%%%%%%%%%%%%%%%%%%%%%%%%%%%%%%%%%%%%%
%%%%%%%%%%%%%%%%%%%%%%%%%%%%%%%%%%%%%%%%%%%%%%%%%%%%%%%%%%%%%%%%%%

\end{document}